# Cell-cycle-synchronized, oscillatory expression of a negatively autoregulated gene in *E. coli*

Zach Hensel and Tatiana T. Marquez-Lago*

Integrative Systems Biology Unit, Okinawa Institute of Science and Technology, Onna-son, Okinawa 904-0495, Japan.

* Correspondence: tatiana.marquez@oist.jp

**Engineering genetic networks to be both *predictable* and *robust* is a key challenge in synthetic biology. Synthetic circuits must reliably function in dynamic, stochastic and heterogeneous environments[1], and simple circuits can be studied to refine complex gene-regulation models[2-5]. Although robust behaviours such as genetic oscillators have been designed and implemented in prokaryotic[6] and eukaryotic[7] organisms, *a priori* genetic engineering of even simple networks remains difficult, and many aspects of cell and molecular biology critical to engineering robust networks are still inadequately characterized. Particularly, periodic processes such as gene doubling and cell division are rarely considered in gene regulatory models, which may become more important as synthetic biologists utilize new tools for chromosome integration[8,9]. We studied a chromosome-integrated, negative-feedback circuit based upon the bacteriophage $\lambda$ transcriptional repressor Cro and observed strong, feedback-dependent oscillations in single-cell time traces. This finding was surprising due to a lack of cooperativity, long delays or fast protein degradation[10]. We further show that oscillations are synchronized to the cell cycle by gene duplication, with phase shifts predictably correlating with estimated gene doubling times. Furthermore, we characterized the influence of negative feedback on the magnitude and dynamics of noise in gene expression. Our results show that cell-cycle effects must be accounted for in accurate, predictive models for even simple gene circuits. Cell-cycle-periodic expression of $\lambda$ Cro also suggests an explanation for cell-size dependence in lysis probability[11] and an evolutionary basis for site-specific $\lambda$ integration[12].**

There are surprisingly few studies of simple negative feedback networks with the single-cell, timelapse resolution required to study gene expression dynamics[6,13-15]. We made a synthetic construct in which the fluorescent fusion protein mVenus-Cro is expressed from the $\lambda$ promoter $P_R$ (**Fig. 1a**). The construct includes the $O_R1$ sequence, which overlaps $P_R$ and binds Cro dimers, repressing transcription[16-18]. This construct was incorporated into the *E. coli* chromosome (replacing the *lac* operon), resulting in strain NF[lac]. In the strain NF[$\Delta cro$,lac], negative feedback is eliminated because only mVenus is expressed from $P_R$ (**Fig. 1b**). NF[lac] and NF[$\Delta cro$,lac] growth was observed for hundreds of minutes; movies showed nucleoid-localized mVenus-Cro fluorescence for NF[lac] (**Fig. 1a**) and cytoplasmic mVenus localization for NF[$\Delta cro$,lac] (**Fig. 1b**). Cro nucleoid localization is consistent with ~50% of Cro molecules being non-specifically bound to DNA at Cro concentrations above ~100 nM[19]. Movies revealed apparent oscillatory expression for NF[lac], which was unexpected for a negatively autoregulated gene lacking cooperative feedback, long transcription/translation delays or fast protein degradation[10,20,21]. Cellular fluorescence levels often appeared to exhibit oscillations with periods approximately equal to one cell cycle. Such an effect was not observed for NF[$\Delta cro$,lac], suggesting that oscillatory expression was not simply an effect of periodic gene doubling, but that oscillations required or were amplified by negative autoregulation. We analysed timelapse fluorescence data for NF[lac] and NF[$\Delta cro$,lac] colonies to estimate the concentration of mVenus fluorophores. Each colony provided ~100 possible trajectories of single-cell mVenus concentration. Typical NF[lac] and NF[$\Delta cro$,lac] mVenus-Cro concentration trajectories (**Figures 1a** and **1b**) show oscillations for NF$^{lac}$, but not for NF$^{\Delta cro,lac}$. Oscillations are less evident when looking at the number of mVenus-Cro molecules per cell rather than concentration (**Supplementary Fig. 1**).

To analyse the apparent, negative-feedback-dependent oscillations, we processed data for entire cell lineages. **Figure 1c** shows a portion of the entire colony lineage corresponding to the timelapse data in **Figure 1a.** Both NF$^{lac}$ and NF$^{\Delta cro,lac}$ exhibit clear correlation in mVenus-Cro concentration time traces that can be seen, for example, by comparing daughter cells (**Supplementary Figs. 2—6** show additional single-cell traces and lineage data for each strain considered in this study, corresponding to **Supplementary Movies 1—5**). NF$^{lac}$ trajectories frequently move above and below the mean expression level, but, in the absence of negative feedback, NF$^{\Delta cro,lac}$ trajectories often take long excursions above or below the mean concentration. For NF$^{lac}$, mVenus-Cro concentration was ~7% that of mVenus in NF$^{\Delta cro,lac}$ without repression (**Supplementary Table 1**). An earlier study found that autoregulated Cro expression from $P_R$ repressed expression from $P_{RM}$ to ~3% its basal level in the absence of Cro[14]. For NF$^{lac}$, correlated expression peaks in daughter cells often occurred over 30 minutes after cell division, and single NF$^{lac}$ time traces frequently exhibited oscillations with periods approximately equal to the cell cycle (**Fig. 1a**, **Supplementary Fig. 2**). We then investigated whether, on average, there was any cell-cycle dependence in mVenus-Cro concentration. **Figure 1d** shows the average mVenus concentration as a function of the fraction of the cell cycle elapsed. In the absence of negative feedback, there is modest cell-cycle dependence. When negative feedback is added, there is a strong and oscillatory cell-cycle dependence with an amplitude of ~10% of the mean mVenus-Cro concentration This cell-cycle-averaged oscillation magnitude is smaller than that observed in individual traces (oscillations in **Fig. 1a** have amplitudes ranging from 25% to 75% of the mean concentration), indicative of noise in cell-cycle length and mVenus-Cro expression.

The observation that oscillations persist when looking at averaged cell cycles rather than single-cell trajectories suggests that an accurate description of mVenus-Cro autoregulation must include parameters that vary in phase with the cell cycle (*i.e.* in synchrony). In other words, some mechanisms could still contribute to oscillations with the same period, but not the same phase, as the cell cycle. So, they would not be sufficient to yield synchrony. Additionally, stochastic discrete effects could lead to oscillations, but would inevitably cause lack of coherence, and averaging over many generations would show no cell-cycle dependence. Possible mechanisms for generating such cell-cycle-synchronized mVenus-Cro oscillations can then be divided into two classes: those that are sensitive to gene location on the chromosome (*e.g.* gene doubling) and those that are not (*e.g.* cell-cycle-periodic changes in global transcription or translation rates). Replication of the *E. coli* chromosome begins from a single origin of replication and oppositely oriented replication forks traverse approximately equal genomic distances before completing replication[22]. Thus, we expected that varying the integration location of the negative feedback construct would induce a phase shift in cell-cycle-synchronized oscillations if synchronization were sensitive to the timing of gene doubling within a cell cycle.

To test whether oscillations were sensitive to the time of gene doubling, the negative-feedback construct was integrated into three alternative genomic loci (**Fig. 2a**). In NF$^{ori}$, the construct is integrated near the origin of replication, while in NF$^{ter1}$ and NF$^{ter2}$ it is integrated near the replication terminus. Accordingly, expected doubling times are 81 (NF$^{ori}$), 59 (NF$^{lac}$ and NF$^{\Delta cro,lac}$), 33 (NF$^{ter1}$) and 37 (NF$^{ter2}$) minutes before cell division[23]. We analysed mVenus-Cro expression time traces for the new strains, and mean mVenus-Cro concentrations varied from 179 (NF$^{ter2}$) to 324 (NF$^{ori}$) molecules/µm$^3$ (**Supplementary Table 1**). Previous experiments exhibited a similar degree of expression level variation when inserting identical constructs at different *E. coli* loci[8,9]. Other than having different mean expression levels, time traces for the new strains (**Fig. 2b**) appeared to also have fewer mid-cell-cycle expression peaks than what was observed for NF$^{lac}$; NF$^{ter2}$ did not exhibit oscillations.

The cross-correlation of the mVenus-Cro concentration time series with a time series consisting of sharp peaks near cell-division times was calculated to measure whether and how oscillations correlated with gene duplication and/or cell division (**Fig. 2c**). Consistent with our observations of averaged cell cycles, NF$^{lac}$ shows stronger cross-correlation than NF$^{\Delta cro,lac}$, with periodic correlation and anti-correlation persisting across ~5 cell generations. NF$^{ori}$ and NF$^{ter1}$ exhibit oscillations that are similarly cell-cycle periodic. The timing of peaks ahead of the cell cycle corresponds to ~20 minutes after approximate gene doubling times, and exhibits precisely the phase shift predicted if gene doubling indeed synchronized mVenus-Cro oscillations. NF$^{ter2}$ shows only weak correlation with the cell cycle. Similar phase shifts were also observed in plots of expression as a fraction of the cell cycle (**Supplementary Fig. 7a**) for NF$^{ori}$ and NF$^{ter1}$. Cross-correlation plots for NF$^{ori}$, NF$^{lac}$, and NF$^{ter1}$ were distinguished not only by a shift in relative phase and magnitude, but also in other ways that were reproduced in independent experiments (**Supplementary Fig. 7b**). The ability to reproduce this complex cell-cycle dependence will be a key test for gene-regulatory modelling of this system.

Lastly, we examined the extent to which negative feedback modulates noise amplitude and bandwidth for our regulatory circuit. **Supplementary Table 1** lists noise measures; for each strain, the coefficient of variation (CV) and Fano factor were calculated from tens of thousands of individual mVenus-Cro concentration time points (concentration distributions shown in **Supplementary Fig. 8a**). The CV is lower in the unregulated case (0.031) than for the other strains (0.063—0.081); this is consistent with studies in which negative feedback increased the CV[3,24]. The Fano factor is higher for unregulated than for autoregulated strains (129 compared to 12.1—24.8 mVenus molecules/µm3). Lower Fano factors are expected with negative feedback, and the range of Fano factors for the autoregulated strains may indicate a range of repression strengths[25-27].

In addition to affecting the magnitude of gene expression noise, feedback has been shown to increase the bandwidth of protein fluctuations by shifting noise to higher frequencies[15,25]. The autocorrelation was calculated from mVenus-Cro time traces (**Fig. 3a**). Results of independent experiments were largely reproducible (**Supplementary Fig. 8b**), with NF$^{\Delta cro,lac}$ always exhibiting slower autocorrelation decay than the strains with negative feedback. The negative feedback strains previously identified as having strong cell cycle dependence (NF$^{ori}$, NF$^{lac}$, and NF$^{ter1}$) also exhibited damped oscillations in the autocorrelation with periods of approximately one cell generation. A previous experiment estimated gene expression noise bandwidth (frequency range) as the reciprocal of the autocorrelation half-time[15]. By this definition, it is clear from **Figure 3a** that negative feedback increases noise bandwidth. However, our larger data sets allow for improved autocorrelation calculations from which the power spectral density (PSD) was estimated for each strain (**Fig. 3b**). The PSDs show that negative feedback increases noise bandwidth by shifting noise to higher frequencies. The PSDs also reveal finer details that further differentiate the oscillating negative feedback strains (NF$^{ori}$, NF$^{lac}$ and NF$^{ter1}$) from the non-oscillating strain with lower expression (NF$^{ter2}$).

In **Figure 4**, we suggest possible molecular mechanisms underlying Cro expression oscillations. First, our Cro circuit may, in the absence of influence from the cell cycle, exhibit damped oscillations of a similar frequency that can be driven by a cell-cycle-periodic oscillation such as gene copy number or concentration. Second, oscillations could be induced by delays between the time when repressors necessarily unbind to allow gene duplication and their subsequent rebinding. Lastly, chromosome conformation changes—particularly positive DNA supercoiling—are known to be associated with DNA replication forks[28]. Supercoiling dependence of $P_R$ transcription and/or Cro binding to $O_R1$ may interact with cell-cycle-dependent supercoiling to create cell-cycle-synchronized Cro oscillations.

Our observations suggest that cell-cycle-dependence may play a critical role in bacteriophage $\lambda$ physiology. For example, it was recently observed that lysis probability increases with cell age[11], while our results suggest increased Cro expression in the second half of the cell cycle. Lamboid prophages integrate at specific sites and have been found to exhibit biases in chromosomal distribution and orientation[12]. Interestingly, we observed the strongest oscillations for strain NF$^{lac}$, in which the construct is inserted within the region of highest prophage density, near the $\lambda$ *attB* site. Our results suggest one mechanism by which chromosome integration sites favouring lysis late in the cell cycle may confer an adaptive advantage for $\lambda$: larger/older cells produce more bacteriophage particles[29]. Given that negative feedback by transcription factors is a common regulatory motif in *E. coli*[30], our results suggest a simple mechanism for cell-cycle-periodic gene expression that may be of wide importance in *E. coli* gene regulation.

**Acknowledgements**

We thank Jie Xiao for providing pJB106 and pZH051 plasmids used as PCR templates for generating strains in this study.

**Author contributions**

The results in this paper are part of a project conceived, designed and supervised by TML. ZH and TML designed experiments. ZH performed experiments and analysed fluorescence images. ZH and TML wrote the manuscript.

**Competing financial interests**

The authors declare no competing financial interests.

**Figures**

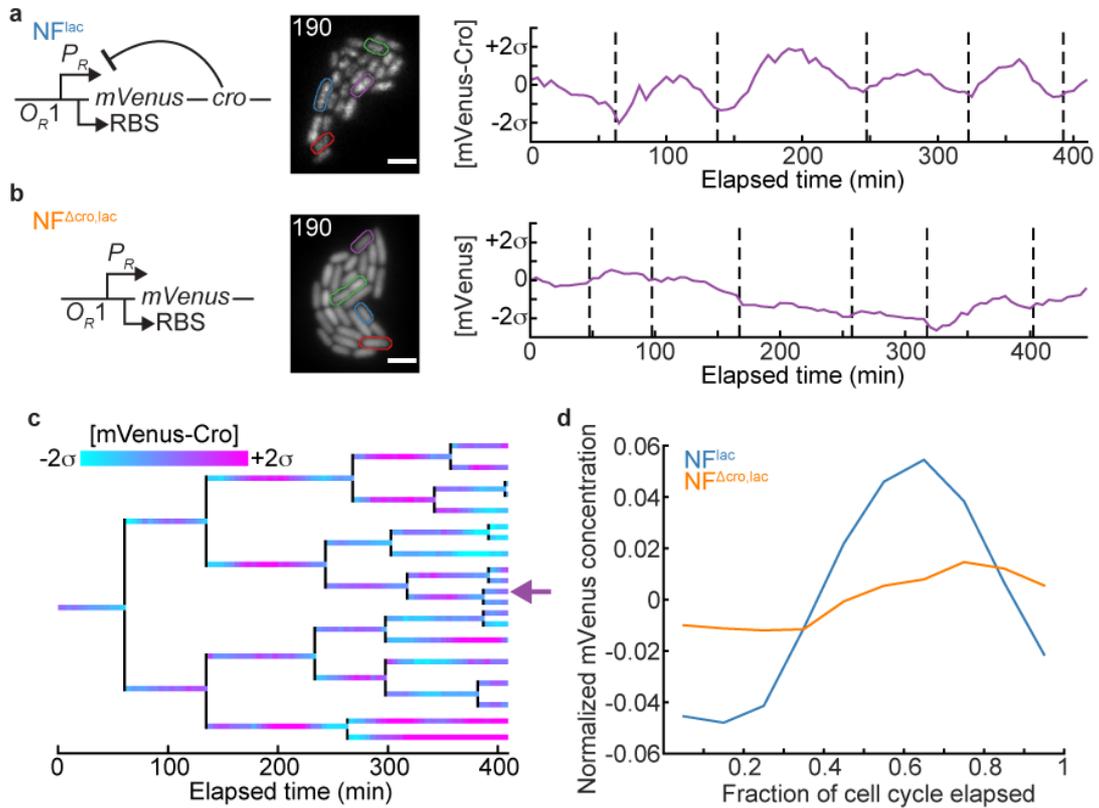

**Figure 1.** Single-cell, timelapse imaging of a simple, negative-feedback gene circuit. (**a**) Left: Cartoon of the NF$^{lac}$ regulatory circuit. Middle: Sample timelapse fluorescence image for NF$^{lac}$ at time 190 minutes from **Supplementary Movie 1**, where cell outline colours indicate cells used for time traces in this figure and **Supplementary Figure 2**. Scale bar is 2 μm. Fluorescence intensity scaled linearly from 0 (black) to 650 (white) mVenus-Cro molecules/μm². Right: single-cell NF$^{lac}$ mVenus-Cro concentration trajectory (dimensionless units, due to scaling by standard deviation). Plot shows mVenus-Cro concentration minus the colony mean concentration and is scaled by the standard deviation of mVenus-Cro concentration. Dashed lines indicate cell division times. (**b**) Left and right: Prepared identically to **a** except for strain NF$^{\Delta cro,lac}$. Middle: Single frame from **Supplementary Movie 2**, where cell outline colours indicate cells used for time traces in this figure and **Supplementary Figure 3**. Fluorescence intensity scaled from 0 (black) to 6,500 (white) mVenus molecules/µm$^2$. (**c**) Concentration of mVenus-Cro for a partial NF$^{lac}$ lineage tree. Each horizontal line corresponds to a single cell generation. Black lines correspond to cell division times and connect mother/daughter cells. A purple arrow indicates the lineage corresponding to the trajectory in **a**. Concentrations are scaled by the standard deviation, σ. (**d**) Concentration of mVenus-Cro normalized by the mean expression level for NF$^{lac}$ and NF$^{\Delta cro, lac}$ was compiled for all cell generations between 65 and 105 minutes long. Concentration values were binned according to the fraction of the cell cycle elapsed.

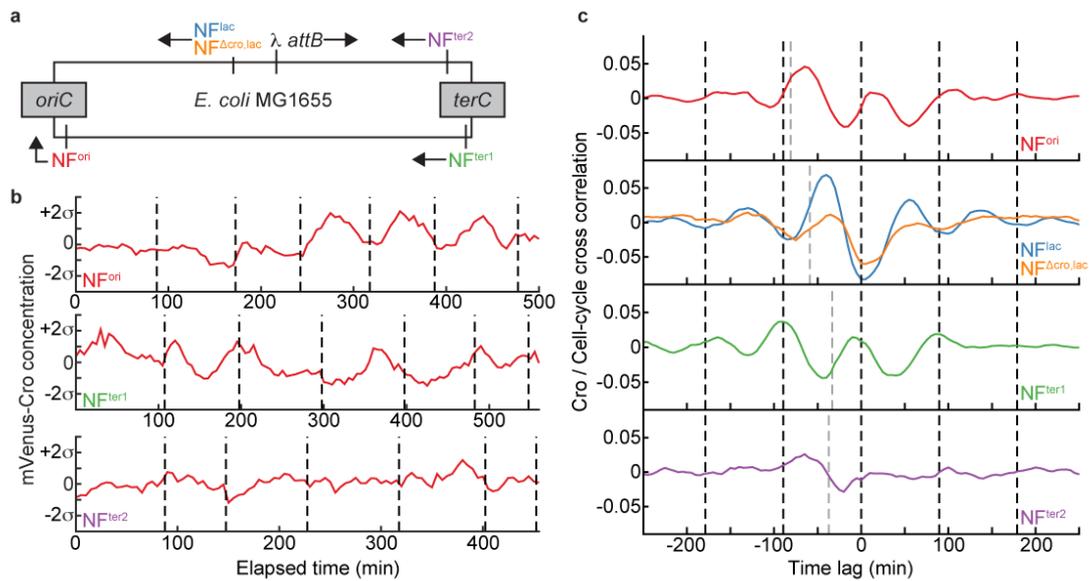

**Figure 2.** Chromosome integration site determines phase shift according to gene doubling time. (**a**) A schematic representation of the *E. coli* MG1655 chromosome (4.6 Mbp). Bidirectional chromosome replication begins from *oriC* (chromosome position 3.9 Mbp) and typically ends near *terC* (1.6 Mbp). Lines and arrows indicate integration sites and $P_R$ transcription directions for NF$^{lac}$/ NF$^{\Delta cro,lac}$ (0.4 Mbp), NF$^{ori}$ (3.9 Mbp), NF$^{ter1}$ (1.6 Mbp) and NF$^{ter2}$ (1.4 Mbp), respectively, as well as the bacteriophage λ *attB* integration site (0.8 Mbp). (**b**) Typical single-cell trajectories for NF$^{ori}$, NF$^{ter1}$ and NF$^{ter2}$ processed identically to those in **Fig. 1**. (**c**) Normalized cross-correlation between the mVenus-Cro concentration and cell-cycle time series, where cross-correlations for all single-cell trajectories were averaged. Peaks at negative/positive time lags indicate that expression is relatively high before/after cell-division times. Dashed black lines indicate average cell division times (89-minute mean generation time). Dashed grey lines indicate estimated time of gene duplication before cell division.

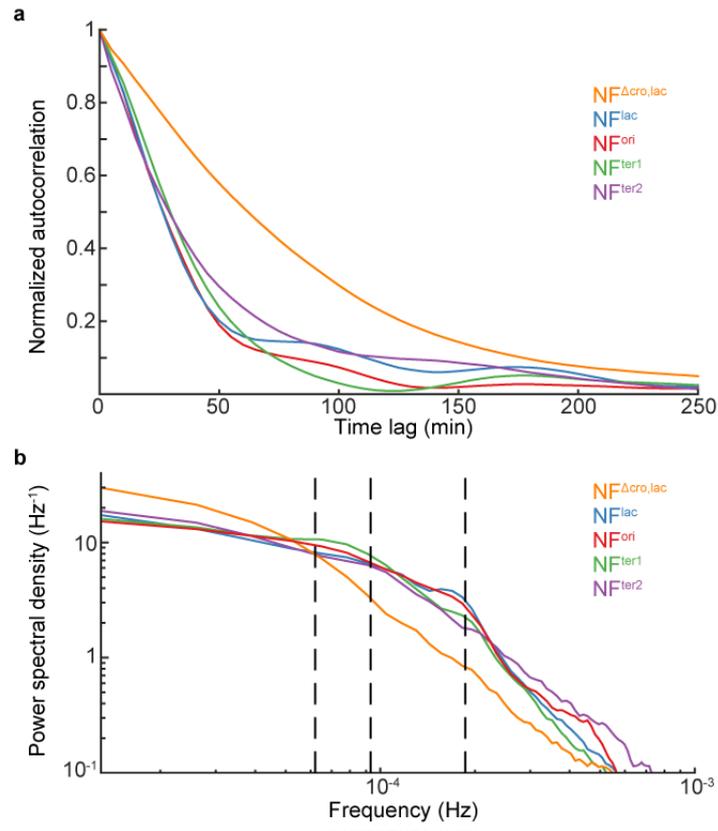

**Figure 3.** Effect of negative feedback on autocorrelation, noise magnitude and frequency. (**a**) The normalized autocorrelation of mVenus-Cro concentration was calculated for every single-cell time trace, and then averaged for each strain. (**b**) The power spectral density is calculated as the FFT of the normalized autocorrelation for each strain. Dashed black lines indicate frequencies associated with 1, 2 and 3 cell generation times.

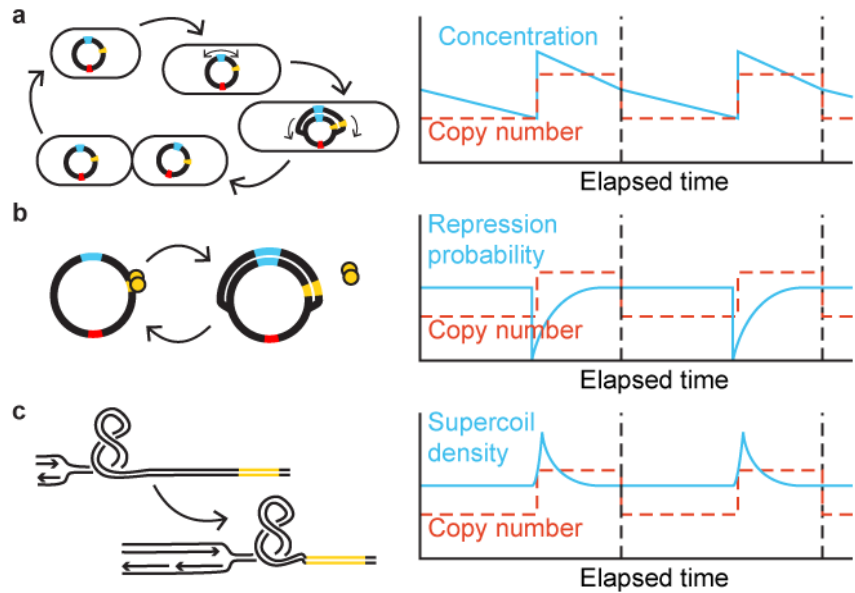

**Figure 4.** Hypothesized mechanisms generating cell-cycle-synchronized oscillations. Each mechanism is illustrated schematically with an associated cell-cycle-periodic parameter than may influence mVenus-Cro oscillations. (**a**) Left, bidirectional replication from *oriC* (blue) to *terC* (red) results in the doubling of the *mVenus-cro* gene (yellow) once per generation, while cell volume monotonically increases until division. Right, hypothetical trajectory of how the copy number and concentration of the *mVenus-Cro* gene would vary in time. Dashed black lines represent cell division times. (**b**) Left, a bound mVenus-Cro dimer (yellow circles) must unbind in order to replicate the *mVenus-cro* gene. Right, this results in a rapid decrease in repression probability that recovers within the timescale in which a mVenus-Cro dimer would re-bind. (**c**) Left, DNA unwinding during chromosome replication causes positive DNA supercoiling; this transiently increases the density of positive supercoils near the *mVenus-cro* gene. Right, local supercoiling near the *mVenus-Cro* gene will become more positive near the time of gene replication and recover within a timescale determined by the activity of enzymes controlling supercoiling homeostasis.

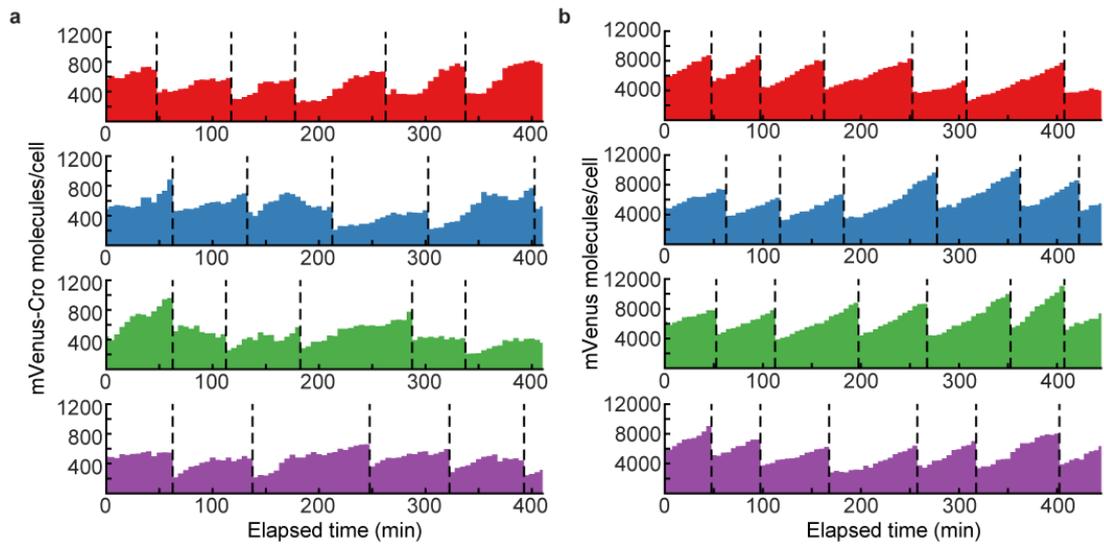

**Supplementary Figure 1.** Raw timelapse data (number of mVenus molecules per cell). (**a**) Data corresponding to NF$^{lac}$ mVenus-Cro concentration time traces in **Figure 1a** and **Supplementary Figure 2a**. Dashed black lines indicate cell division times. (**b**) Data corresponding to NF$^{\Delta cro,lac}$ mVenus concentration time traces in **Figure 1b** and **Supplementary Figure 3a**.

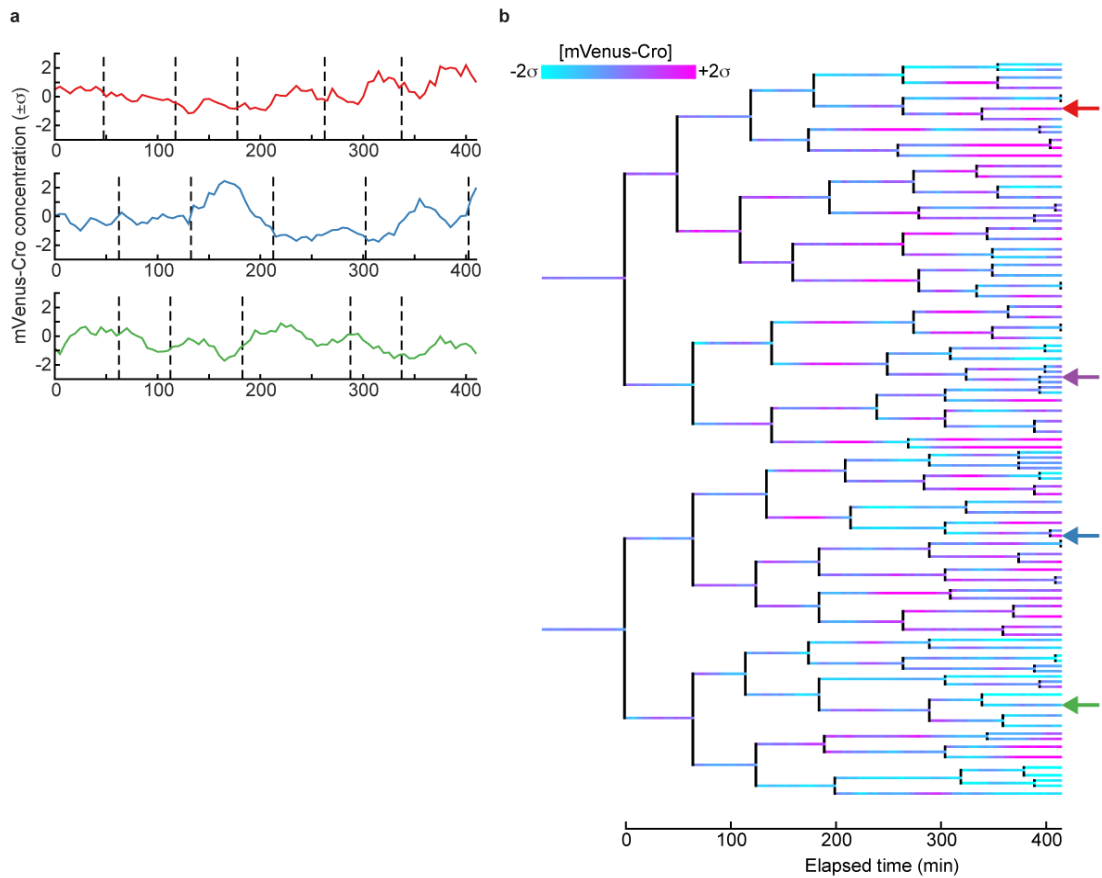

**Supplementary Figure 2.** Additional NF$^{lac}$ timelapse data. (**a**) Three additional single-cell mVenus-Cro concentration trajectories complement the one shown in **Figure 1a**. Dashed black lines indicate cell division times. (**b**) Complete NF$^{lac}$ mVenus-Cro concentration lineage. Concentration values are for the concentration minus the mean colony concentration and are scaled by the standard deviation of NF$^{lac}$ mVenus-Cro concentration according to the inset scale bar. Coloured arrows indicate the final cell in corresponding single-cell trajectories in **Figure 1a** and in this figure, and outlined cells in **Supplementary Movie 1**. The 0-minute time point is aligned to the first frame in which there are 4 cells to facilitate comparison with the corresponding data, but the entire lineage was used in data analysis.

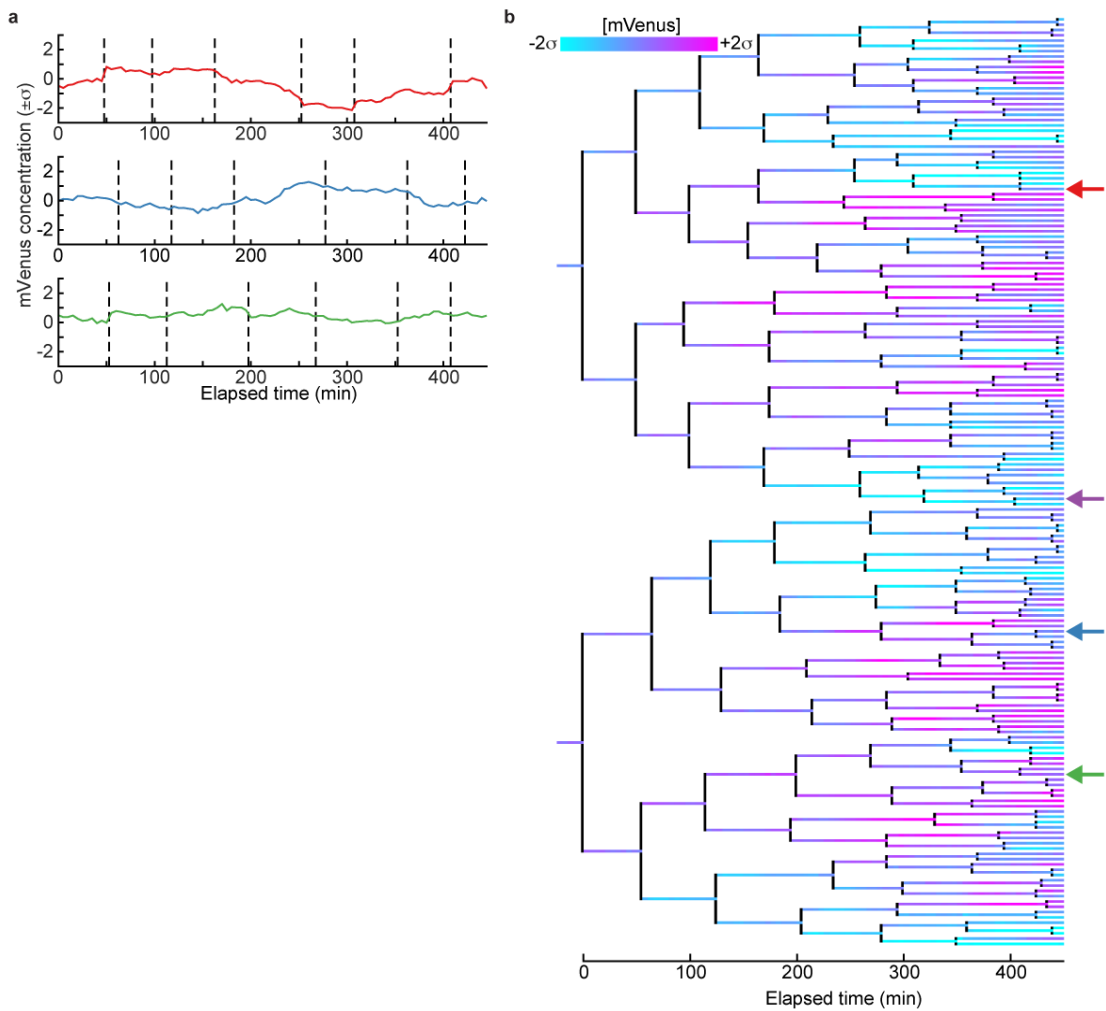

**Supplementary Figure 3.** Additional NF$^{\Delta cro,lac}$ timelapse data. (**a**) Three additional single-cell mVenus concentration trajectories complement the one shown in **Figure 1b**. Dashed black lines indicate cell division times. (**b**) Complete NF$^{\Delta cro,lac}$ mVenus concentration lineage. Concentration values are for the concentration minus the mean colony concentration and are scaled by the standard deviation of NF$^{\Delta cro,lac}$ mVenus concentration according to the inset scale bar. Coloured arrows indicate the final cell in corresponding single-cell trajectories in **Figure 1b** and in this figure, and outlined cells in **Supplementary Movie 2**. The 0-minute time point is aligned to the first frame in which there are 4 cells to facilitate comparison with the corresponding data, but the entire lineage was used in data analysis.

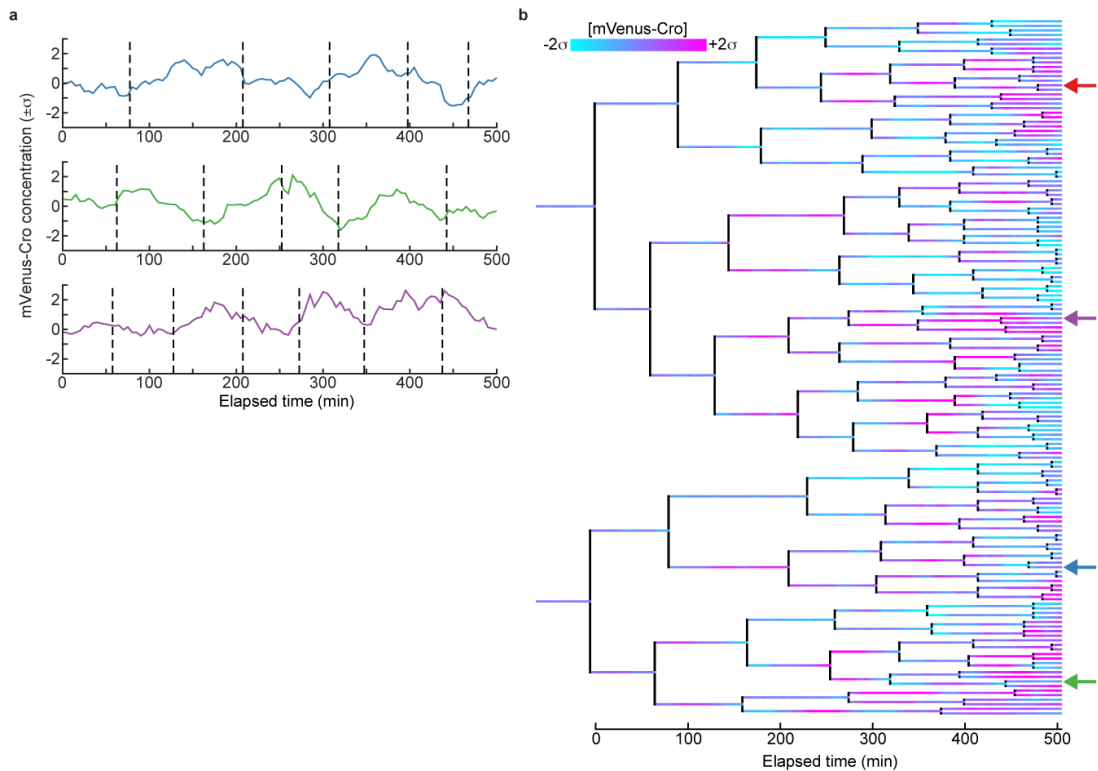

**Supplementary Figure 4.** Additional NF[ori] timelapse data. (**a**) Three additional single-cell mVenus-Cro concentration trajectories complement the one shown in **Figure 2b**. Dashed black lines indicate cell division times. (**b**) Complete NF[ori] mVenus-Cro concentration lineage. Concentration values are for the concentration minus the mean colony concentration and are scaled by the standard deviation of NF[ori] mVenus-Cro concentration according to the inset scale bar. Coloured arrows indicate the final cell in corresponding single-cell trajectories in **Figure 2b** and in this figure, and outlined cells in **Supplementary Movie 3**. The 0-minute time point is aligned to the first frame in which there are 4 cells to facilitate comparison with the corresponding data, but the entire lineage was used in data analysis.

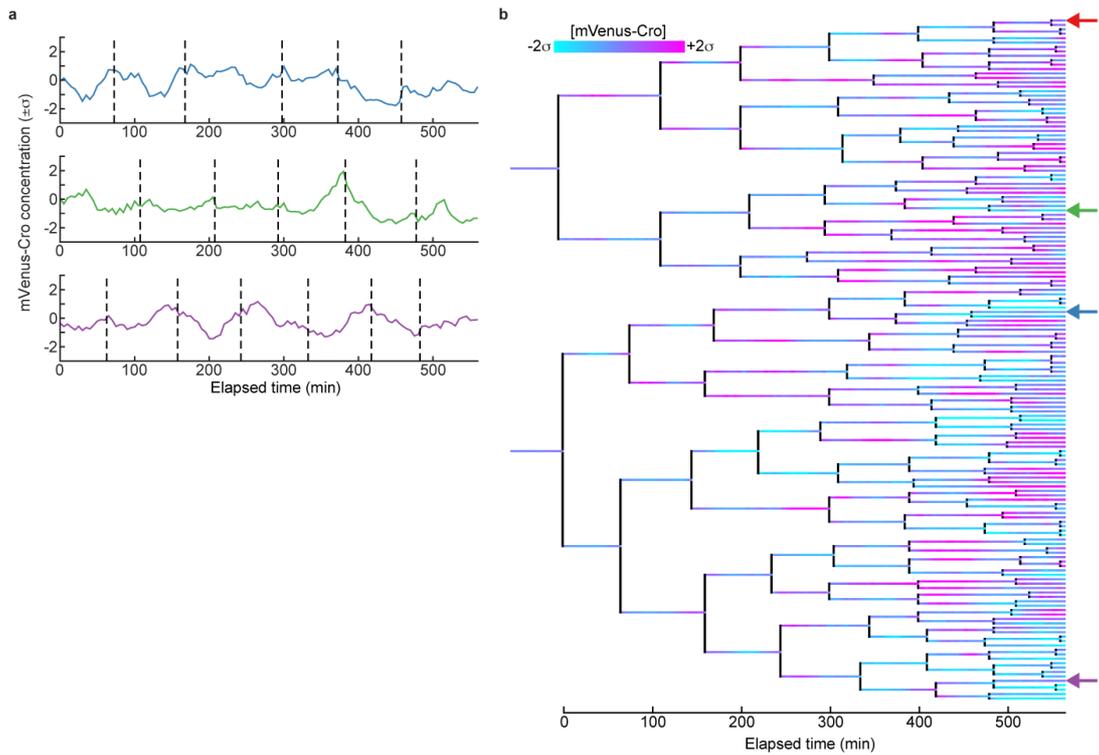

**Supplementary Figure 5.** Additional NF[ter1] timelapse data. (**a**) Three additional single-cell mVenus-Cro concentration trajectories complement the one shown in **Figure 2b**. Dashed black lines indicate cell division times. (**b**) Complete NF[ter1] mVenus-Cro concentration lineage. Concentration values are for the concentration minus the mean colony concentration and are scaled by the standard deviation of NF[ter1] mVenus-Cro concentration according to the inset scale bar. Coloured arrows indicate the final cell in corresponding single-cell trajectories in **Figure 2b** and in this figure, and outlined cells in **Supplementary Movie 4**. The 0-minute time point is aligned to the first frame in which there are 4 cells to facilitate comparison with the corresponding data, but the entire lineage was used in data analysis.

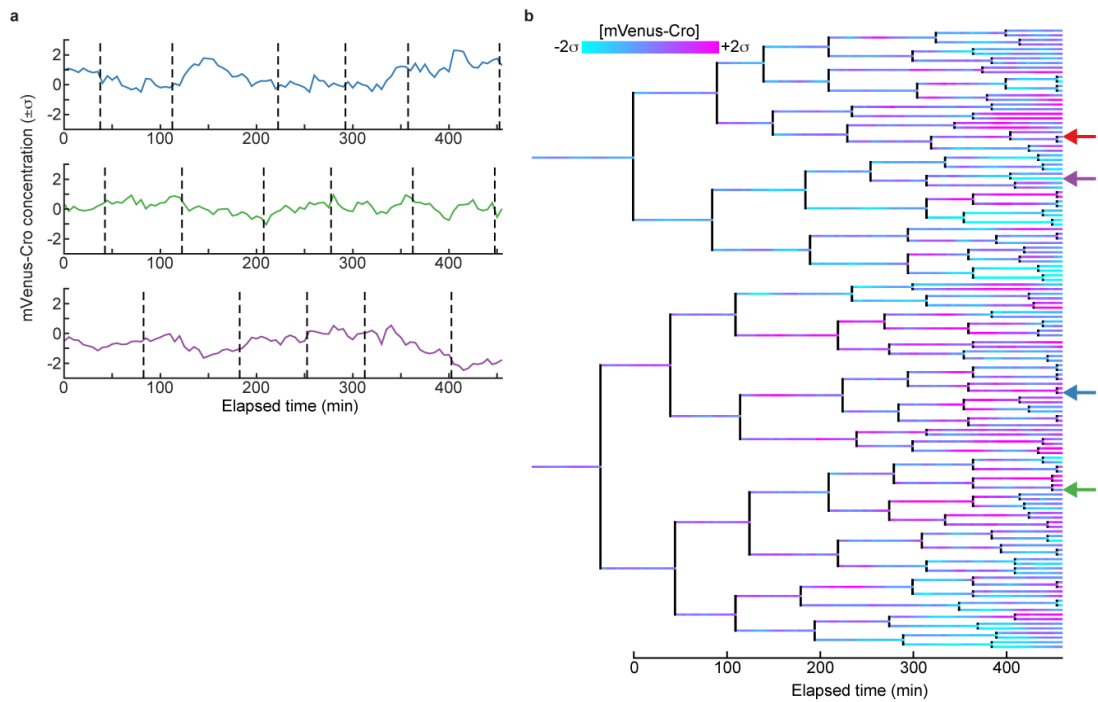

**Supplementary Figure 6.** Additional NF[ter2] timelapse data. (**a**) Three additional single-cell mVenus-Cro concentration trajectories complement the one shown in **Figure 2b**. Dashed black lines indicate cell division times. (**b**) Complete NF[ter2] mVenus-Cro concentration lineage. Concentration values are for the concentration minus the mean colony concentration and are scaled by the standard deviation of NF[ter2] mVenus-Cro concentration according to the inset scale bar. Coloured arrows indicate the final cell in corresponding single-cell trajectories in **Figure 2b** and in this figure, and outlined cells in **Supplementary Movie 5**. The 0-minute time point is aligned to the first frame in which there are 4 cells to facilitate comparison with the corresponding data, but the entire lineage was used in data analysis.

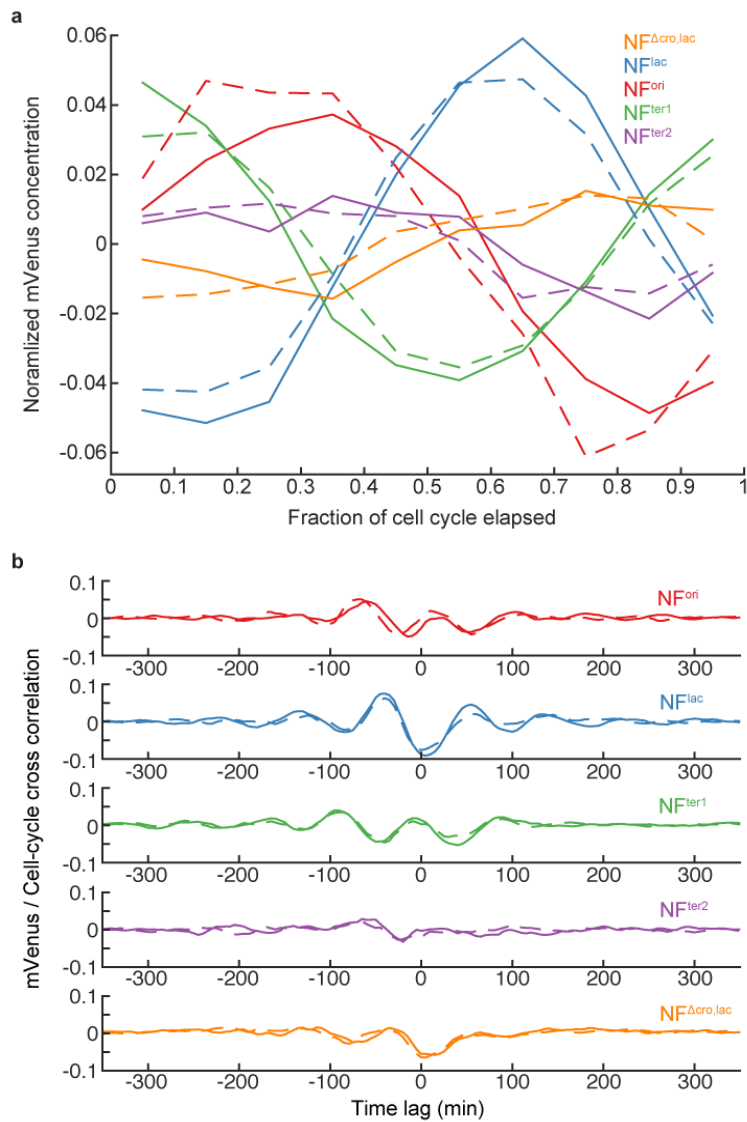

**Supplementary Figure 7.** Reproducibility in cell-cycle-dependent oscillations. (**a**) Normalized mVenus concentration (subtracting the colony mean concentration and normalized by the mean concentration for each strain) is plotted as a function of the fraction of cell cycle elapsed as in **Figure 1d**. Here, all strains are included with data from the first (solid lines) and second (dashed lines) independent experiments. (**b**) Cross correlation between mVenus concentration and the cell cycle is plotted as in **Figure 2c**. Here, data is plotted separately for the first (solid lines) and second (dashed lines) independent experiments.

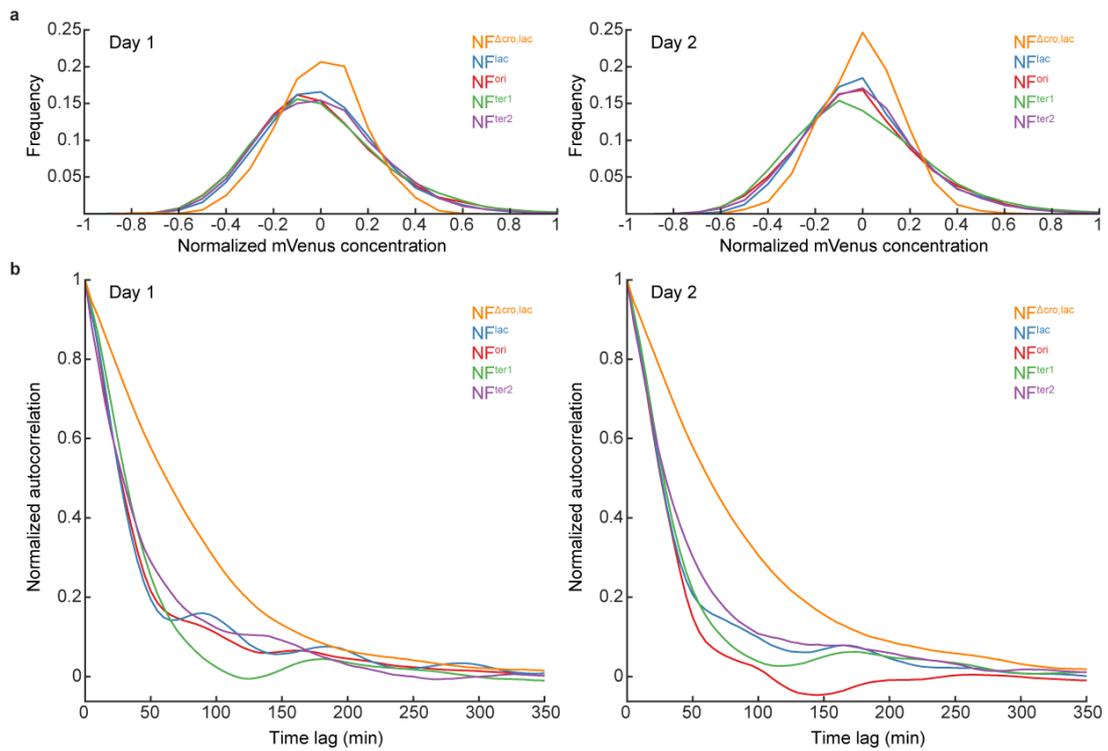

**Supplementary Figure 8.** Reproducibility in mVenus concentration distributions and autocorrelation. (**a**) Histograms of normalized mVenus concentration are plotted (single-cell concentration minus mean colony concentration and normalized by the mean concentration for each strain). Here, data is plotted separately for the first (left) and second (right) independent experiments. (**b**) Average, normalized autocorrelation of mVenus concentration trajectories is plotted as in **Figure 3a**. Here, data is analysed and plotted separately for the first (left) and second (right) independent experiments.

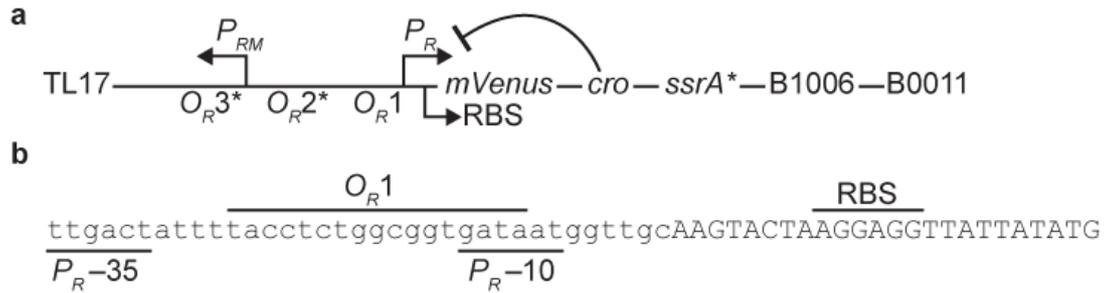

**Supplementary Figure 9.** Components of the NF$^{lac}$ gene circuit. (**a**) The gene *mVenus-cro-ssrA* is expressed from the $P_R$ promoter and encodes a dimeric, autoregulatory protein that binds at $O_R1$ to prevent transcription from $P_R$. The C-terminal SsrA tag carries a mutation that does not lead to fast degradation. The construct also contains a ribosome binding site (RBS), weak promoter $P_{RM}$, transcription terminators (TL17, B1006 and B0011) and two additional $O_R$ sites that have been mutated to reduce Cro binding. (**b**) Relevant regulatory sequences in all constructs showing $P_R$ –35 and –10 consensus sequences and RBS. Nucleotides in the $P_R$ transcript are capitalized and the sequence ends with the start-codon ATG.

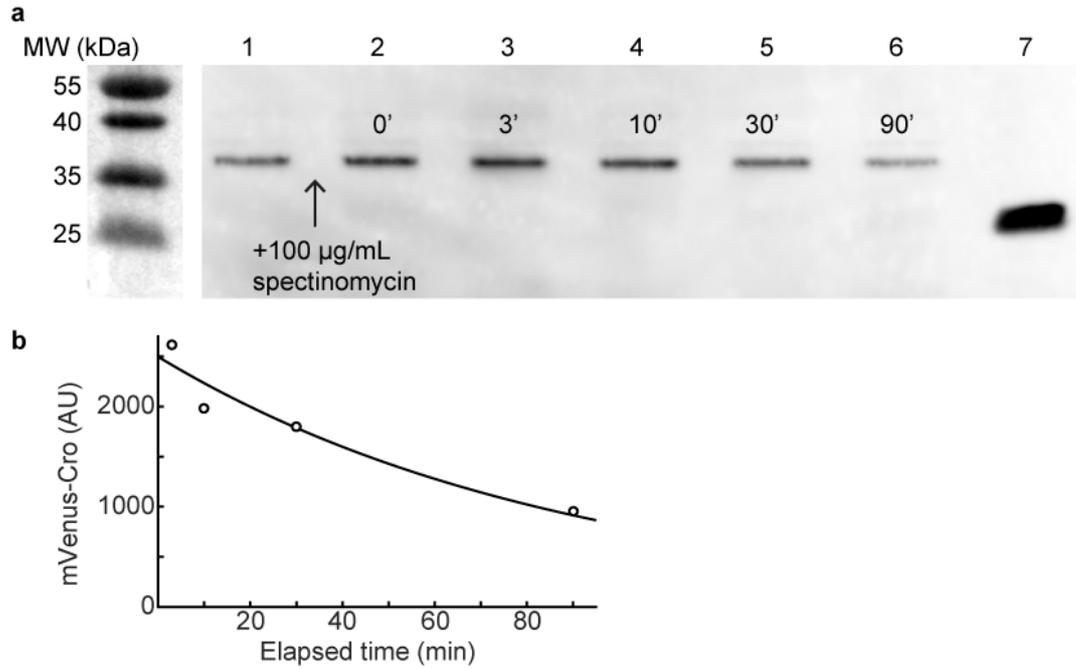

**Supplementary Figure 10.** Mutated SsrA tag does not lead to fast mVenus-Cro degradation. (**a**) Anti-GFP blot for cells harbouring pISB0021 before (lane 1) and at various time points after (lanes 2–6) the addition of spectinomycin to 100 μg/ml. Calculated mVenus-Cro-SsrA molecular weight is 35.2 kDa. Lane 7 contains 100 ng recombinant GFP (~27 kDa). (**b**) Band intensity for lanes 3–6 fit to exponential decay ($k$ = 0.0112 min$^{-1}$; $t_{1/2}$ = 62 minutes) to estimate mVenus-Cro degradation rate with mutated SsrA tag.

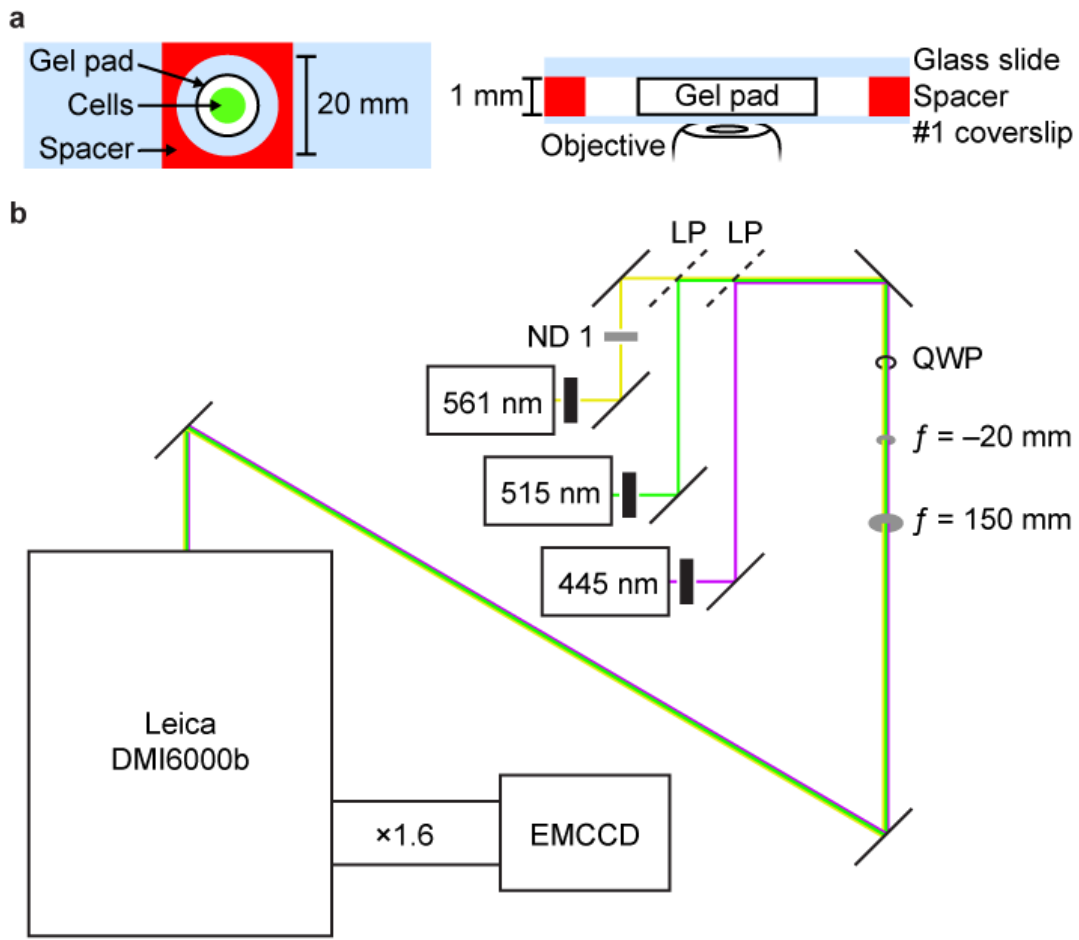

**Supplementary Figure 11.** Microscope sample preparation. (**a**) Microscope slide preparation. Left: 100 μl of agarose is pipetted to the centre of a well formed by a silicon isolator on a coverslide, and the sample is sealed with a coverglass to allow the gel pad to set. The coverglass is then removed, and a small drop of cell suspension is added to the centre of the gel pad and allowed to dry. Right: The sample is then sealed with a #1 coverslip and imaged on an inverted microscope within a temperature-controlled chamber. (**b**) Microscope excitation optics schematic. Three lasers are free-space coupled to the back port of an inverted microscope. Software-controlled shutters (black rectangles) control laser output. A neutral-density (ND) filter reduced the 561-nm laser power. Beams are combined by long-pass filters (dashed lines) and steered by mirrors (solid lines). Beams are circularly polarized by a quarter-wave plate (QWP) and expanded by two lenses in a Galilean configuration (grey ellipses). The output image is additionally magnified by a factor of 1.6 and captured on an EM-CCD.

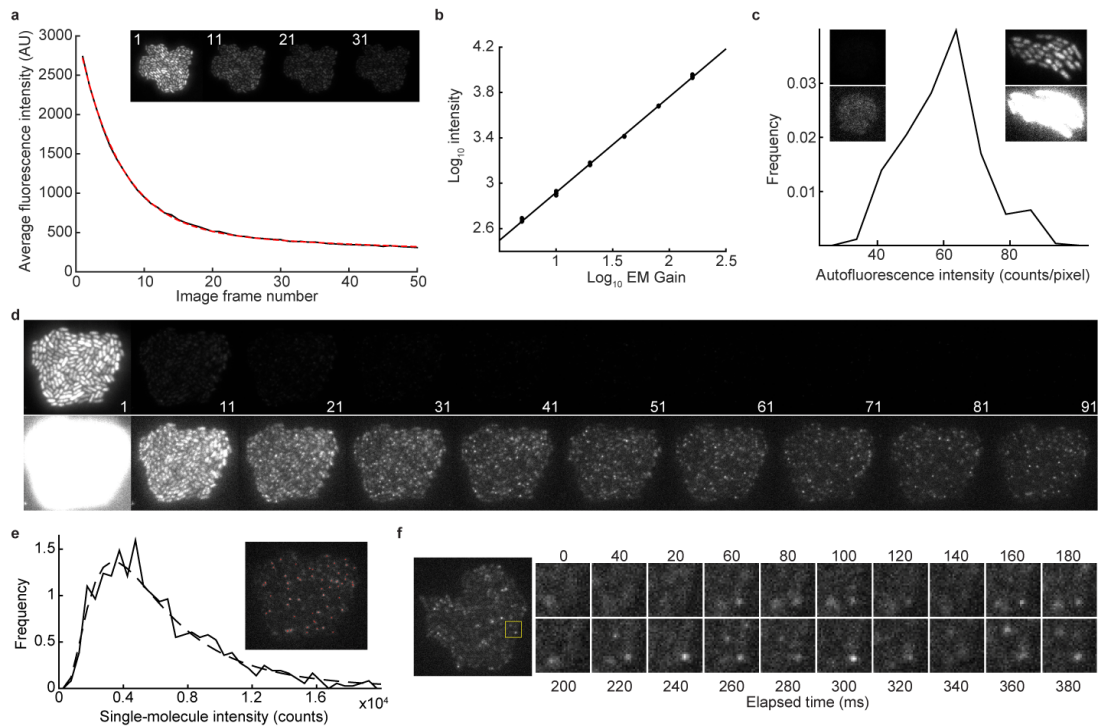

**Supplementary Figure 12.** Quantification and control experiments for mVenus fluorescence. (**a**) A NF$^{ori}$ colony was imaged at 1-s intervals with identical acquisition settings to those used for timelapse imaging. The mVenus photobleaching rate was determined by measuring the average fluorescence of the colony minus the average fluorescence of pixels outside the colony (black line) and fitting as the sum of two exponential decays (dashed red line). Inset: Sample images for frames 1, 11, 21 and 31. (**b**) A microcolony was imaged with varying EM-gain settings (first increasing and then decreasing EM gain, *G*, and imaging 5 times at each step) and the average fluorescence intensity was measured (black dots). After correcting for a 500.6-count offset and ~2% photobleaching per frame, the intensity was found to be proportional to $G^{0.84}$ (solid line). (**c**) The average mVenus-Cro signal per pixel, resulting from autofluorescence, was calculated for single cells from the strain ISB0024 in which pISB0063 expresses mCherry, but mVenus is mutated to be non-fluorescent; a histogram of average intensities is shown. Inset: Left, two images

of ISB0024/pISB0063 displayed with low (top) and high (bottom) maximum intensities. Right, two images for the lowest-expression-level strain, NF$^{ter2}$, acquired with the same imaging conditions and scaled to the same intensities as in the ISB0024/pISB0063 images. (**d**) Sample frames from an NF$^{ori}$ photobleaching experiment show single mVenus-Cro molecules non-specifically bound to the chromosome after tens of acquisitions with 40-mW laser excitation at 1-s intervals. The same images are shown for every 10 image frames with high (top) and low (bottom) maximum intensities. (**e**) Solid line: integrated single-molecule intensity distribution for mVenus-Cro from frames 91—100 of the movie shown in **d**. Dashed line: Fit with lognormal distribution. Inset: red dots show single-molecules identified in frame 91. (**f**) A different NF$^{ori}$ microcolony was imaged using streaming (50 Hz) image acquisition using the same excitation intensity used for **d**. Left: single image of entire colony at 380 ms. Right: zoom view of 20 sequential images within the yellow region-of-interest showing single-molecule fluorescence lasting for several frames.

**Supplementary Movie 1.** Sample microscope data for one NF$^{lac}$ colony. Left: mCherry images used for cell segmentation (arbitrary intensity scaling). Right: mVenus-Cro images with intensities scaled from 0 to 650 mVenus-Cro molecules/μm$^2$ according to the gradient scale bar. Solid scale bar, 5 μm. Colours of outlined cells correspond to single-cell trajectories in **Figure 1a** and arrows in **Supplementary Figure 2**.

**Supplementary Movie 2.** Sample microscope data for one NF$^{\Delta cro,lac}$ colony. Left: mCherry images used for cell segmentation (arbitrary intensity scaling). Right:

mVenus images with intensities scaled from 0 to 6,500 mVenus molecules/µm$^2$ according to the gradient scale bar. Solid scale bar, 5 µm. Colours of outlined cells correspond to single-cell trajectories in **Figure 1b** and arrows in **Supplementary Figure 3**.

**Supplementary Movie 3.** Sample microscope data for one NF$^{ori}$ colony. Left: mCherry images used for cell segmentation (arbitrary intensity scaling). Right: mVenus-Cro images with intensities scaled from 0 to 650 mVenus-Cro molecules/µm$^2$ according to the gradient scale bar. Solid scale bar, 5 µm. Colours of outlined cells correspond to single-cell trajectories in **Figure 2b** and **Supplementary Figure 4**.

**Supplementary Movie 4.** Sample microscope data for one NF$^{ter1}$ colony. Left: mCherry images used for cell segmentation (arbitrary intensity scaling). Right: mVenus-Cro images with intensities scaled from 0 to 650 mVenus-Cro molecules/µm$^2$ according to the gradient scale bar. Solid scale bar, 5 µm. Colours of outlined cells correspond to single-cell trajectories in **Figure 2b** and **Supplementary Figure 5**.

**Supplementary Movie 5.** Sample microscope data for one NF$^{ter2}$ colony. Left: mCherry images used for cell segmentation (arbitrary intensity scaling). Right: mVenus-Cro images with intensities scaled from 0 to 650 mVenus-Cro molecules/µm$^2$ according to the gradient scale bar. Solid scale bar, 5 µm. Colours of outlined cells correspond to single-cell trajectories in **Figure 2b** and **Supplementary Figure 6**.

# Tables

**Supplementary Table 1. Strain descriptions and statistics**

| Strain | cro | Loci (Mbp) | $\mu$ | $\sigma^2/\mu$ | $\sigma^2/\mu^2$ | N |
|---|---|---|---|---|---|---|
| NF$^{\Delta cro,lac}$ | − | 0.4 | 4,031 | 127 | 0.0314 | 25 |
| NF$^{lac}$ | + | 0.4 | 263 | 16.6 | 0.0630 | 29 |
| NF$^{ori}$ | + | 3.9 | 324 | 24.8 | 0.0765 | 30 |
| NF$^{ter1}$ | + | 1.6 | 251 | 20.3 | 0.0808 | 29 |
| NF$^{ter2}$ | + | 1.4 | 179 | 12.1 | 0.0673 | 26 |

Recombination loci are mapped according to the MG1655 genome (Accession #NC_000913.3). The origin of replication, *oriC*, is at 3.9 Mbp and *terC* is at 1.6 Mbp relative to the 4.6-Mbp chromosome. Noise figures of merit were calculated from the standard deviation of mVenus concentration (single-cell concentration minus mean colony concentration), $\sigma$, and the mean mVenus concentration, $\mu$ (mVenus molecules/µm³). The Fano factor is $\sigma^2/\mu$ (mVenus molecules/µm³) and the unitless coefficient of variation (CV) is $\sigma^2/\mu^2$. *N* lists the number of microcolonies imaged for each strain; data is combined from two independent experiments for all strains.

**Supplementary Table 2. Strains used in this study.**

| Nickname | Strain | Plasmid template | Insert DNA primers | CPCR primers |
|---|---|---|---|---|
| NF$^{\Delta cro,lac}$ | ISB0026/pISB0063 | pISB0029 | 1.28<br>1.29 | 1.26<br>1.27 |
| NF$^{lac}$ | ISB0006/pISB0063 | pISB0015 | 1.28<br>1.29 | 1.26<br>1.27 |
| NF$^{ori}$ | ISB0028/pISB0063 | pISB0015 | 2.41<br>2.42 | 2.43<br>2.44 |
| NF$^{ter1}$ | ISB0055/pISB0063 | pISB0015 | 3.16<br>3.17 | 3.18<br>3.19 |
| NF$^{ter2}$ | ISB0054/pISB0063 | pISB0015 | 3.16<br>3.17 | 3.18<br>3.19 |
|  | ISB0024/pISB0063 | pISB0043 | 1.28<br>1.29 | 1.26<br>1.27 |
|  | Top10/pISB0021 | N/A | N/A | N/A |

Strain nicknames correspond to descriptive names used in the main text. Strains harbour different plasmids and are listed as "chromosome name/plasmid name." Insert DNA primers are used for amplifying DNA for homologous recombination. CPCR primers are external to the homology regions used for recombination and are used to amplify the entire recombined insert by colony PCR for sequencing. All strains/plasmids are first reported in this work except *E. coli* Top10 (Invitrogen).

**Supplementary Table 3. Primer sequences for chromosome recombination.**

| Primer | Sequence |
| --- | --- |
| 1.28 | cgcggtatggcatgatagcgcccggaagagagtcaattcagCGGGGCTGGCTTAACTATGC |
| 1.29 | gaacgccagccgccacgacgtttggtggaatgtcttttgtgacGCTATGACCATGCTCGAGCC |
| 1.26 | ccatcgaatggcgcaaaacc |
| 1.27 | cgcgaataacccgacaagg |
| 2.41 | ccagtgcatacaattgcgacttttctgctaaccctgttcgaCGGGGCTGGCTTAACTATGC |
| 2.42 | taaatatgctgtgcgcgaacatgcgcaatatgtgatctgaagGCTATGACCATGCTCGAGCC |
| 2.43 | cgcccccgtgatttcaaac |
| 2.44 | gcaaaggcatcatttgccaag |
| 3.16 | caactattcagatacatcactcccatcacattcattcctccgCGGGGCTGGCTTAACTATGC |
| 3.17 | gaagcggcaactgcaaactatcttatgtagagactctacacggGCTATGACCATGCTCGAGCC |
| 3.18 | ccaatcgcagcacgttcttg |
| 3.19 | gaatgctaacaggtggcagc |

Sequences of primers used for chromosome recombination and colony PCR verification when constructing the strains listed in **Supplementary Table 2**. For chromosome recombination primers, sequences with homology to the MG1655 genome are lowercase and homology to the synthetic construct is uppercase.

**Supplementary Table 4. Statistics from independent experiments.**

|  | | Generation length | | | mVenus concentration | | |
|---|---|---|---|---|---|---|---|
|  | | | | |  | $\sigma$ | |
|  | Day | $N$ | $\mu$ | $\sigma$ | $\mu$ | Mean-subtracted | Raw |
| NF$^{\Delta cro,\text{lac}}$ | 1 | 849 | 86.9 | 26.5 | 3,682 | 682 | 819 |
|  | 2 | 786 | 92.5 | 31.7 | 4,386 | 747 | 1,253 |
| NF$^{\text{lac}}$ | 1 | 1,299 | 93.5 | 29.1 | 257 | 64.7 | 78.7 |
|  | 2 | 852 | 92.2 | 31.2 | 263 | 68.2 | 77.0 |
| NF$^{\text{ori}}$ | 1 | 1,746 | 87.2 | 29.1 | 327 | 90.5 | 98.3 |
|  | 2 | 728 | 85.9 | 28.1 | 319 | 87.7 | 98.2 |
| NF$^{\text{ter1}}$ | 1 | 1,778 | 91.2 | 30.3 | 254 | 71.5 | 76.1 |
|  | 2 | 1,288 | 90.4 | 27.5 | 251 | 71.2 | 77.0 |
| NF$^{\text{ter2}}$ | 1 | 969 | 82.4 | 25.3 | 167 | 43.6 | 46.7 |
|  | 2 | 1,253 | 90.5 | 29.0 | 179 | 48.5 | 53.0 |

Experiment days correspond to plots in **Supplementary Figures 7** and **8**. The number of complete generations, $N$, generation length (minutes) mean, $\mu$, and standard deviation, $\sigma$, are listed for each independent experiment. For mVenus concentration (molecules/µm³), the mean, $\mu$, and standard deviation of concentration, $\sigma$, are listed for each experiment. The standard deviation is shown for both the raw mVenus concentration timelapse data and for data in which the mean mVenus fluorescence of the colony was subtracted from each data point. Mean-subtracted standard deviations were used for calculating all noise statistics and for scaling all figures showing mVenus concentration timelapse data (single-cell time traces and cell lineages).

**Online Methods**

***E. coli* strain development**

Our synthetic circuit was based on the bacteriophage lambda genomic sequence. An initial construct was obtained by synthesis in a plasmid (Genewiz). The synthetic plasmid included: the lambda $O_R$ operator ($P_{RM}$ through $P_R$), a codon-optimized mVenus-Cro-SsrA ORF ending in a double stop codon, and three bidirectional transcription terminators[31-33] (TL17 and BioBricks parts B1006 and B0011) as shown in **Supplementary Figure 9a**. The YFP mVenus is useful in gene-expression studies owing to its rapid chromophore maturation[34]. The sequence of the 11-amino-acid, C-terminal SsrA tag was mutated from its wild-type sequence[35] (AANDENYALAA) to a mutated sequence (AANDENYALAD) that is not recognized by the ClpXP protease[36].

Preliminary experiments indicated mVenus-Cro levels were substantially lower than the expected ~1000 molecules/cell, so RBS-strengthening mutations were designed and tested using an online calculator[37], resulting in the stronger RBS shown in **Supplementary Figure 9b**. Both $O_R2$ and $O_R3$ were mutated in the pISB0015 plasmid to bind Cro more weakly. Mutated sequences are "taacaccgtgcCCgttg" and "tatcaGcAcaaTggata" with capital letters indicating mutated bases that confer reduced binding strengths of ~4 kcal/mol[38]. In our constructs, Cro negatively regulates its own expression by binding the higher affinity, wild-type $O_R1$ sequence, so binding to the mutated $O_R2$ and $O_R3$ sites is negligible (note also that Cro binding is only weakly cooperative[21]). The relatively weak $P_{RM}$ promoter is present, but terminates after only 29 bases and contains no RBS. A kanamycin resistance cassette was amplified from pZH051[14] for selection in chromosome integration. The no-fluorescence construct ISB0024 was made by first introducing the G67A mutation to mVenus in pISB0026 to mutate the chromophore, resulting in plasmid pISB0043 which was used as a template for constructing ISB0024.

A no-feedback construct was constructed by removing mVenus-SsrA from the ORF to make pISB0029. The mutant SsrA sequence was not expected to result in rapid degradation, but mVenus-Cro-SsrA likely differs from Cro to some extent with respect to dimerization and DNA-binding affinities as well as protein folding and degradation rates. We attempted to make two alternative "no-feedback" constructs. First, we introduced the Cro K56T mutation reported to reduce Cro-DNA binding while minimally reducing Cro stability[39]. However, preliminary experiments showed that strains bearing this mutation did not have increased mVenus-Cro expression expected from reducing negative feedback. Second, we attempted to make mutants with extremely weak-binding $O_R1$ mutations, but repeated cloning attempts were unsuccessful; we attributed this to toxic effects when mVenus-Cro is overexpressed and binds non-specifically to DNA.

The plasmid pISB0021 was used for degradation rate measurements and to verify that the mutated SsrA tag did not lead to rapid mVenus-Cro degradation (**Supplementary Figure 10**). It includes the original, weaker, RBS and the wild-type $O_R2$ sequence; neither of which is expected to influence mVenus-Cro degradation. To measure the degradation rate, the *E. coli* Top10 cells harbouring pISB0021 were grown to OD600 = 0.4 at 30° C in a 10-mL culture. A 0.5-mL sample was taken before adding spectinomycin to 100 µg/ml to inhibit translation of new mVenus-Cro. Then, 0.5-ml samples were taken after continued incubation (0, 3, 10, 30 and 90 minute samples). All samples were immediately pelleted, resuspended in 100 µL loading buffer (NEB B7703), and kept on ice. A recombinant GFP standard (100 ng; Clontech 632373) was similarly prepared and all samples were boiled at 97° C for 10 minutes, briefly centrifuged, and analysed by SDS-PAGE along with a visible ladder (Bio-Rad 162-0177). Proteins were transferred to a PVDF membrane, blocked overnight (4° C, 1% BSA in TBST), and detected with anti-GFP primary (Abcam ab1218, 1:2,000 in blocking buffer), Goat-anti-mouse-HRP secondary (Bio-rad 171-1011, 1:25,000 in TBST) and HRP signal (Takara Western BloT Hyper HRP) was imaged (FujiFilm LAS 3000). Band intensities were quantified using NIH ImageJ.

The pISB0063 plasmid, in which mCherry is expressed from the IPTG-inducible T5-lac promoter, was made by replacing the ORF in pJB106[40] with a codon-optimized mCherry ORF (synthesized by Genewiz). Standard molecular biology techniques were used to construct all plasmids (restriction/ligation, site-directed mutagenesis by linear and inverse PCR, and blunt-end ligation). Codon-optimization consisted of random selection based on weighted probabilities of codon usage for highly expressed *E. coli* genes[41].

Plasmid-borne constructs were inserted into the *E. coli* chromosome by homologous recombination. Strain MG1655 harbouring plasmid pKD46[42] was transformed by electroporation using linear DNA amplified from pISB0015, pISB0029 or pISB0043, and recombinants were selected for kanamycin resistance. The pISB0063 plasmid was transformed into recombinant strains by the one-step TSS method[43]. Complete strain and plasmid details are listed in **Supplementary Table 2**. Primer sequences used for targeting to different chromosomal loci are listed in **Supplementary Table 3**. Colony PCR using primers inside and outside of the synthetic sequence screened chromosome recombinants, and colony PCR products were completely sequenced. Note that the same primers were used for NF$^{ter1}$ and NF$^{ter2}$ because the targeted sequences have extremely high homology; both NF$^{ter1}$ and NF$^{ter2}$ colonies were obtained from the same transformation, and strain identity was determined by sequencing.

**Microscopy**

Cells were grown overnight from fresh colonies (LB-kanamycin plates) in M9 minimal media supplemented with 1× MEM amino acids, 0.4% glucose, 20 µM IPTG, 50 µg/ml kanamycin and 35 µg/ml chloramphenicol at 30° C with shaking. Several dilutions of culture were made so that one would be at an appropriate density (OD600 ~0.05—0.2) to begin microscopy after ~16 h, without disrupting exponential growth. Cells were diluted (OD600 ~0.04) and a 0.5-µl sample was spotted onto ~100-µl gel pads (growth media plus 3% SeaPlaque GTG agarose; Lonza), allowed to dry for ~5 minutes at room temperature, and microscope samples were assembled as illustrated in **Supplementary Figure 11a** and rapidly moved to the microscope which was preheated to 30° C.

Cells were imaged on a temperature-controlled Leica DMI6000B inverted microscope equipped with a 100× oil-immersion objective, additional 1.6× magnification, and an EM-CCD (Photometrics Evolve 512) with 16-µm pixels for a final pixel size of 100 nm. Fluorescence images were acquired with 5-mW laser excitation from a Coherent Sapphire 514-nm laser and 1-mW excitation from a Coherent Sapphire 561-nm laser (10-mW beam through neutral density filter). The beams were circularly polarized and expanded (7.5× using −20-mm and 150-mm focal-length achromatic doublet lenses in a Galilean configuration) before entering the back port of the microscope. Unless otherwise noted, optical components were purchased from ThorLabs. Thus, ~0.7-mm diameter laser beams were expanded to 5.25 mm, resulting in a ~52.5-µm spot after 100× magnification. **Supplementary Figure 11b** is a diagram of the microscope setup. All components were controlled using Metamorph software (Molecular Devices).

In each experiment, 15 different locations were imaged at 5-minute intervals starting from one or two cells and continuing for hundreds of minutes (experiments ranged from 330 to 630 minutes with a median of 525 minutes). For each acquisition the focal plane was acquired using the microscope's built-in interferometric focusing device. The colony outline was then identified and the stage was moved to centre the colony. Brightfield, mCherry (custom filter set with Semrock filters FF01-561/13, FF02-616/73 and DI02-R561), and mVenus (Semrock LF514-B filter set) images were acquired. **Supplementary Movies 1— 5** show typical mCherry and mVenus movies for each strain. To minimize excitation power variation, a 30-µM square region (300×300 pixels) was imaged, with cell colonies rarely extending past the central ~25-µM (250×250 pixels).

Data for some colonies was discarded due to failures in the automated image protocol or poor focus, and movies were not analysed after microcolonies overgrew the imaging area grew out of the focal plane. Cells exhibited generation times of 89 ± 29 minutes (N=11,548 generations; mean ± 1 S.D.). Cell growth rates varied between colonies for a single experiment, but generation-length distributions were consistent between different strains and for separate experiments (**Supplementary Table 4**).

**Image analysis**

Cell colonies were segmented into single cells using semi-automated processing of mCherry images largely following a previously described method[14]. Our methods differed from earlier work in that mCherry images were used rather than brightfield images, which dramatically improved the fidelity of automatic segmentation. Segmentation was then manually checked and corrected using a custom MATLAB script. Objects were morphologically thickened, images were aligned based upon the centre of mass of detected cells, and cells were assigned to cell lineages based on maximum overlap of cells in adjacent time frames. Automatic assignment occasionally failed in the instance of large shifts in colony morphology; assignments were manually checked and corrected by visual inspection of the segmentation data.

For mVenus data, single-molecule imaging was used to convert fluorescence counts into number of fluorescent proteins, while mCherry data was kept in arbitrary fluorescence units. The integrated intensity in arbitrary fluorescence units was estimated by taking the total intensity of pixels corresponding to cells in segmented mCherry images and subtracting the median intensity of pixels outside of any cells. We acquired mVenus data every 1 s using identical settings to those used for timelapse imaging in order to estimate the rate of mVenus photobleaching (**Supplementary Fig. 12a**). We then measured the integrated intensity of an NF$^{ori}$ colony minus the intensity of pixels outside the colony. The average intensity of pixels within colony was well fit by the function:

$$I = 2.6 \times 10^3 e^{-0.166t} + 0.5 \times 10^3 e^{-0.0092t}$$

Thus, most photobleaching is attributed to a single exponential decay process with a 20.8-minute half-life. In other words, ~15% of mVenus molecules were photobleached at each acquisition time; over the same 5 minutes, in the absence of new protein production, the mVenus-Cro concentration decreases by ~9% (combining the measured rates of protein degradation rate, 0.011 min$^{-1}$, and dilution by cell growth, 0.0077 min$^{-1}$).

Timelapse data for NF$^{\Delta cro,lac}$ was acquired with a different EM Gain setting (20) than that used for other strains (200) because of higher expression levels for NF$^{\Delta cro,lac}$. We acquired fluorescence images using the same amplification and digitization settings across a range of EM Gain values to determine the conversion factor, 6.918, needed to convert NF$^{\Delta cro,lac}$ fluorescence intensities into values comparable to those for other strains (**Supplementary Fig. 12b**). Autofluorescence background was estimated from images of the strain ISB0024/pISB0063 where non-fluorescent mVenusG67A-Cro is expressed. ISB0024/pISB0063 cells were segmented as described above and found to have an apparent mVenus background of 60.06 counts/pixel (**Supplementary Fig.**

**12c**). This background was subtracted from all mVenus fluorescence data, giving a reduction of ~10 mVenus molecules/μm$^3$.

In order to convert integrated fluorescence intensities from arbitrary fluorescence units to an approximate number of mVenus molecules, we measured the distribution of fluorescence intensities of single mVenus molecules in the NF$^{ori}$ strain. Images were acquired at 1-s intervals for 100 s (using the same growth and imaging conditions used for timelapse image acquisition, except for a laser power of 40 mW rather than 5 mW). The movie initially showed nucleoid-localized mVenus fluorescence, with later images showing diffraction-limited, single-molecule spots (**Supplementary Fig. 12d**). Using the ThundersSTORM[44] plug-in in FIJI[45], spots were identified in the final 10 frames and fit to a Gaussian distribution to estimate single-molecule intensities. The spot intensity distribution was well fit by a lognormal distribution with a mean of 6,171 counts (**Supplementary Fig. 12e**); a lognormal distribution is expected for a distribution of spots at different focal planes[46]. We note that single mVenus spots are observed to discretely blink on and off and generally persist for more than one frame (**Supplementary Fig. 12f**). We then divided integrated cellular fluorescence intensities by 771 (the mean single-molecule intensity divided by 8 to account for the change from 40- to 5-mW excitation) to estimate the number of fluorescent mVenus molecules per cell.

Cell area in pixels was measured from the manually corrected segmentation images. A small correction factor of 20 pixels was added to each cell (relative to a median area of ~800 pixels), as we observed a discontinuity in the mVenus/cell-cycle cross-correlation at $\tau = 0$ that was inconsistent with the assumption that, on average, there is no abrupt discontinuity in mVenus concentration at the time of cell division. The area correction approximately corresponds to the median cell diameter, which is intuitive since the morphological thickening operation described above expands cells until they are separated by ~1 pixel. So the 20-pixel correction roughly corresponds to the border between recently divided cells, which becomes less significant as a fraction of cell area as the cell cycle progresses. Cell volume, $V$, was estimated from the projected cell area, $A$, from a simple rectangular approximation to a cylinder:

$$A = 2rL$$

$$V = \pi r^2 L = \pi r A / 2$$

Here, $r$ is the median cell radius (9.8 pixels in the segmented image, or 490 nm) and $L$ is the cell length (inferred from area and median radius). The median radius was estimated as one half the minor axis length of all segmented cells (using the bwregionprops function in MATLAB). Units were converted from pixels$^3$ to µm$^3$ based on the optical magnification (160×), pixel size (16 µm) and 2× image resizing during image segmentation. The mean cell volume was 1.62 µm$^3$, which is smaller than the 3-µm$^3$ volume previously reported for growth conditions that were similar (MG1655, M9 media with glucose), but at 37° rather than 30° C[47].

Representative movies (**Supplementary Movies 1—5**) were generated in MATLAB with mVenus intensities scaled according to the estimated number of molecules per µm$^2$ (following the same background subtraction and quantization procedures described above). Fluorescence images were first aligned using the image segmentation results, and cell outlines were generated from image segmentation data. Sample frames from these movies (**Fig. 1a** and **1b**) were also generated in MATLAB; individual frames were extracted using ImageJ and assembled in Adobe Illustrator.

**Timelapse data analysis**

Timelapse data consisted of the estimate number of mVenus-Cro molecules and cell volume plus information required to construct single-cell lineages (parent/daughter cell identities and cell division times). Following earlier work[15], the mVenus-Cro "concentration noise" was analysed (single-cell concentration minus the average concentration for all cells in one colony). The only time the raw concentration data was used was for calculating the mean concentration. Using raw concentration data rather than concentration noise gives broadly similar results to those we report, but gives more day-to-day variation of replicate experiments (**Supplementary Table 4**). We hypothesize that this is because a sample size of no more than 15 microcolonies is too small to completely sample long-lived cell-to-cell variations in cell growth rates and other parameters. In order to estimate mVenus concentration statistics (**Supplementary Table 1**) and distributions (**Supplementary Fig. 8**), every data point was used from every complete cell generation, combining data from two days' independent experiments (Total N≥1,635 generations for each strain; **Supplementary Table 4**). Importantly, statistical distributions can differ if they are gathered from many samples at a single time point (population statistics) or a single sample at multiple timepoints[3]. Differences depend upon whether the measured process is ergodic, initial conditions, and how quickly the system evolves relative to sampling frequency and duration. In this experiment, sampling is a hybrid of these regimes, with several hundred data points (i.e. from each colony cell lineage) for each initial condition (i.e. for each strain, data was collected from 25—30 separate colonies).

Cell-cycle-averaged expression levels (**Fig. 1d**, **Supplementary Fig. 7a**) were calculated by combining all mVenus concentration data for complete cell cycles between 65 and 105-minutes long (the middle 3 quintiles of the generation-length distribution). For a cell cycle of length $t_{gen}$, the time for each frame ($t$ = 5, 10, … tgen minutes) was used to estimate the fractional time:

$$t_{frac} = t/t_{gen}$$

Data pairs of mVenus concentration and fractional time were binned according to fractional time (bins = 0.05, 0.10, … 0.95) and the average concentration was plotted for each bin, normalized by the strain's mean expression level. Our conclusions were generally insensitive to which generations were included, but including more generations results in weaker apparent cell-cycle dependence, and including fewer generations results in insufficient sampling.

Cross-correlation between mVenus concentration and the cell cycle (**Fig. 2c**, **Supplementary Fig. 7b**) was calculated for every possible single-cell trajectory for every movie. The biased cross-correlation was calculated as:

$$[f(t) \star g(t)](\tau) = \frac{1}{N\sigma_f\sigma_g} \sum_{n=0}^{N-\tau-1} (f(n+\tau) - \langle f \rangle)(g(n) - \langle g \rangle)$$

Here, $\tau$ is the time lag, $t$ is elapsed time, $N$ is the trajectory length, $\sigma$ is the standard deviation of each time series, brackets denote averages over a single trajectory and $f$ is the mVenus concentration trajectory. The cell-cycle trajectory $g$ is created by convolving a trajectory equalling one at cell division times and zero elsewhere with a sharp Gaussian distribution ($\sigma$ = 0.75 frames or 3.75 minutes), to approximate the error in determining the true cell division time. We note that whether or not this was done made no difference in the conclusions we drew from this analysis. The average cross-correlation for each time lag was calculated by averaging over all trajectories including data for that time lag.

The normalized, biased autocorrelation (**Fig. 3a**, **Supplementary Fig. 8b**) was calculated following a previously reported method[15] for each single-cell trajectory of length $N$:

$$[f(t) \star f(t)](\tau) = R(\tau) = \frac{1}{N\langle f^2 \rangle} \sum_{n=0}^{N-\tau-1} f(n+\tau)f(n)$$

Symbols are the same as those used for cross-correlation, as is the procedure used to average over all time lags. Note that $R(\tau) = R(-\tau)$ and $R(0) = 1$ follow from the definition of autocorrelation for finite time series used in this study. There were significant differences in autocorrelation magnitude for some replicate experiments (especially for NF[ori]), indicating insufficient sampling to precisely determine autocorrelation magnitudes while sampling a small number of colonies. The power spectral density (PSD) was calculated as the fast Fourier transform of the autocorrelation using MATLAB's fft function (N=256; sample sizes ranged from 209 to 251, from maximum autocorrelation time lags ranging from 520 to 625 minutes). Truncating autocorrelation data to the same length (*e.g.* from 0 to 500-minute time lags) did not significantly affect the PSD. This PSD-calculation method follows the Wiener-Khinchin theorem[48], which holds if the mean and autocorrelation of a process is time invariant.

Both autocorrelation and cross-correlation use biased estimation methods (not correcting for sample size and, by definition, approaching zero at large, poorly sampled time lags). To ease comparison, we followed here methods established by an earlier study using similar data[15]. We also note that there are several possible ways to estimate autocorrelation from cell lineage data, all of which carry benefits and drawbacks. We used all possible trajectories from a given lineage. This oversamples earlier cell generations, which appear in many trajectories relative to the latest cell generations that are only in single trajectories. One alternative method is to solely use unique autocorrelation pairs, but this oversamples later time points for the colony since more cells exist at that point and will biased if there is any global time-dependence in the experiment. We also implemented this method, but it did not appreciably affect any of our conclusions or PSD calculations. Lastly, one could sample only one trajectory from each colony, but this makes it difficult to obtain a sufficiently large data set and would require choosing samples through quasi-arbitrary criteria.

**Estimated gene doubling times**

Replications of the circular, 4.64-megabase *E. coli* chromosome begins from a single origin of replication, *oriC*, and oppositely oriented replication forks traverse approximately equal genomic distances before completing replication, usually near the *terC*/*dif* locus[22]. Complete chromosome replication for *E. coli*

MG1655 grown takes ~42—54 minutes followed by a 33—34 minute delay before cell division in similar growth conditions to those in our study[23]. Expected doubling times assumed 48 minutes for chromosome replication at a constant speed and a 33-minute delay before division are were calculated from the fraction of each chromosome half between *oriC* and each recombination site. For example, NF$^{lac}$ is inserted at 0.36 Mbp, 1.26 Mbp before *terC*, so NF$^{lac}$ replication occurs at approximately (each half of the chromosome is 2.32-Mbp long):

$$33' + 48' \times \frac{1.26}{2.32} = 59'$$

Gene duplication times for all strains were estimated in this way.

**Reproducibility**

Timelapse experiments for all strains were repeated in two independent experiments (summaries in **Supplementary Table 4** and **Supplementary Figs. 7** and **8**). Analysis in the main text and **Supplementary Table 1** uses data compiled from both independent experiments.

**Figure preparation**

All plots were generated in MATLAB and figures were assembled using Adobe Illustrator. Linear intensity scaling was used for all images and movies.